\documentclass[12pt]{article}
\usepackage{feynmf}
\newcommand{\fmftil}[1]{\fmfv{de.sh=circle,de.fi=0,de.si=2thick}{#1}}
\newcommand{\fmfgam}[2]{\fmfv{d.sh=circle,d.fi=-.5,d.si=#1}{#2}}
\title{
{\vspace{-3cm} \normalsize \hfill
                        \parbox{33mm}{MS-TPI-96-4 \\
                                      hep-th/9603145}  }\\[25mm]
Tadpole Summation by Dyson-Schwinger Equations}

\author{Jens K\"uster and Gernot M\"unster \\
        Institut f\"ur Theoretische Physik I,
        Universit\"at M\"unster\\
        Wilhelm-Klemm-Str.~9, D-48149 M\"unster, Germany}
\date{March 16, 1996}
\begin{document}
\maketitle

\begin{abstract}
In quantum field theory with three-point and four-point couplings the
Feynman diagrams of perturbation theory contain momentum independent
subdiagrams, the ``tadpoles'' and ``snails''.
With the help of Dyson-Schwinger equations we show how these can be
summed up completely by a suitable modification of the mass and coupling
parameters.
This reduces the number of diagrams significantly.
The method is useful for the organisation of perturbative calculations
in higher orders.
\end{abstract}
%
\section{Introduction}
In quantum field theoretical perturbation theory it is sometimes
desirable or even indispensable to sum up infinite classes of Feynman
diagrams.
For example, in the framework of field theory at finite temperatures
characteristic properties of some propagators only reveal themselves
after certain resummations of the perturbation series have been
performed \cite{Kapusta}.
Another case are perturbative calculations beyond the two-loop level,
where resumming subclasses of diagrams can be very helpful in taming
the large number of diagrams and their combinatorics.

In the more interesting cases the classes of diagrams to be summed up
are nested in some way.
This can for example result from an iterative definition of these
diagrams.
The correct counting of diagrams is then not trivial and their
resummation may be susceptible to errors.
The literature contains examples for this kind of problems; for a
discussion see e.g.\ \cite{BBH}.
Therefore it is advantageous to rely on systematic methods when classes
of diagrams are to be summed.

In this article we discuss the summation of tadpole (and related)
diagrams in a quantum field theory with three-point and four-point
couplings.
With the help of Dyson-Schwinger equations all tadpole subdiagrams, to
be specified below, can be summed up.
This leads to modified Feynman rules such that the set of Feynman
diagrams, which need to be considered, reduces to those without
tadpoles.
In order to be definite we restrict ourselves to the case of a scalar
field theory.
The method, however, is essentially of combinatoric nature and can be
generalized to other models straightforwardly.

Our motivation comes from a three-loop calculation in scalar field
theory in the phase with broken symmetry \cite{GKM}.
The number of diagrams we encountered is already nonnegligible: there
are 204 one-particle irreducible diagrams contributing to the inverse
propagator at the three-loop level, compared to 20 at two loops.
After exploiting symmetries there are still 162 of them.
Many of them contain tadpoles and their elimination by means of
Dyson-Schwinger equations simplifies the book-keeping very much.
The reduced set of one-particle irreducible propagator-diagrams without
tadpoles only contains 34 elements.
At the four-loop level the gain is even larger.

The term tadpole diagram (in the context of Feynman diagrams) is used
differently in the literature.
An early reference is \cite{Salam}.
We use the following notation.
The ``bare tadpole'' is the following subdiagram
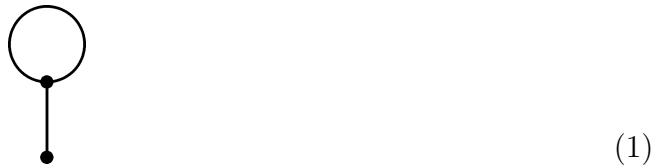
\begin{fmffile}{mftadpole}
\begin{equation}
    \mbox{\begin{fmfchar}(10,20)
        \fmfincoming{i1,i2,i3}
        \fmfoutgoing{o1,o2,o3}
        \fmfdot{v1,v2}
        \fmf{plain,tension=0}{v1,v2}
        \fmf{plain,tension=0,left}{v2,v3,v2}
        \fmf{phantom}{i1,v1,o1}
        \fmf{phantom}{i2,v2,o2}
        \fmf{phantom}{i3,v3,o3}
      \end{fmfchar}}
\end{equation}
which is attached with its tail to the rest of a Feynman diagram.
Obviously it exists only in theories with  three-point vertices.
A ``tadpole diagram'' is a connected subdiagram which is connected with
the rest of a Feynman diagram only through a single line, like these
examples:
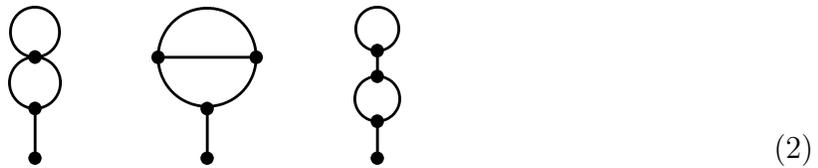
\begin{equation}
    \mbox{\begin{fmfchar}(20,20)
        \fmfincoming{i1,i2,i3,i4}
        \fmfoutgoing{o1,o2,o3,o4}
        \fmfdot{v1,v2,v3}
        \fmf{plain,tension=0}{v1,v2}
        \fmf{plain,tension=0,left}{v2,v3,v4,v3,v2}
        \fmf{phantom}{i1,v1,o1}
        \fmf{phantom}{i2,v2,o2}
        \fmf{phantom}{i3,v3,o3}
        \fmf{phantom}{i4,v4,o4}
      \end{fmfchar}
      \begin{fmfchar}(20,20)
        \fmfincoming{i1,i2,i3,i4}
        \fmfoutgoing{o1,o2,o3,o4}
        \fmfdot{v1,v2,v3,v4}
        \fmf{plain,tension=0}{v1,v2}
        \fmf{plain,tension=0,left}{v3,v4,v3}
        \fmf{plain,tension=0}{v3,v4}
        \fmf{phantom}{i1,v1,o1}
        \fmf{phantom}{i2,v2,o2}
        \fmf{phantom}{i3,v3,v5,v6,v7,v4,o3}
      \end{fmfchar}
      \begin{fmfchar}(20,20)
        \fmfincoming{i1,i2,i3,i4,i5}
        \fmfoutgoing{o1,o2,o3,o4,o5}
        \fmfdot{v1,v2,v3,v4}
        \fmf{plain,tension=0}{v1,v2}
        \fmf{plain,tension=0.5}{v3,v4}
        \fmf{plain,tension=0,left}{v2,v3,v2}
        \fmf{plain,tension=0,left}{v4,v5,v4}
        \fmf{phantom}{i1,v1,o1}
        \fmf{phantom}{i2,v2,o2}
        \fmf{phantom}{i3,v3,o3}
        \fmf{phantom}{i4,v4,o4}
        \fmf{phantom}{i5,v5,o5}
      \end{fmfchar}}
\end{equation}
There is no momentum flow from a tadpole diagram to the rest of the
diagram.
Normal ordering would remove the bare tadpole but not all of the tadpole
diagrams.
The sum of all tadpole diagrams, graphically denoted by
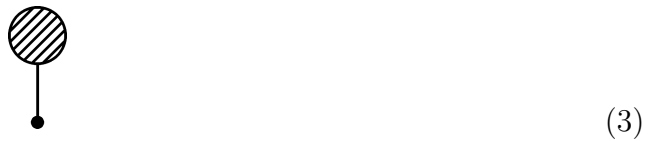
\begin{equation}
    \mbox{\begin{fmfchar}(10,15)
        \fmfincoming{i1,i2}
        \fmfoutgoing{o1,o2}
        \fmfblob{.5h}{v2}
        \fmfdot{v1}
        \fmf{plain,tension=.2}{v1,v2}
        \fmf{phantom}{i1,v1,o1}
        \fmf{phantom,tension=.5}{i2,v2,o2}
      \end{fmfchar}}
\end{equation}
is called the ``tadpole''.

Apart from the tadpole diagrams there is another class of subdiagrams
which do not have momentum flowing in or out.
These are the ``bare snail'' (also known as the slug),
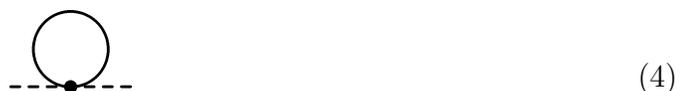
\begin{equation}
    \mbox{\begin{fmfchar}(20,10)
        \fmfincoming{i1,i2}
        \fmfoutgoing{o1,o2}
        \fmfdot{v1}
        \fmf{plain,tension=0,left}{v1,v2,v1}
        \fmf{dashes}{i1,v1}
        \fmf{dashes}{o1,v1}
        \fmf{phantom}{i2,v2,o2}
      \end{fmfchar}}
\end{equation}
and the ``snail diagrams'', which are connected with the rest of a
Feynman diagram through two lines which meet at a vertex, like these
examples
\begin{equation}
  \mbox{\begin{fmfchar}(20,20)
      \fmfincoming{i1,i2,i3}
      \fmfoutgoing{o1,o2,o3}
      \fmfdot{v1,v2}
      \fmf{plain,tension=0,left}{v1,v2,v3,v2,v1}
      \fmf{dashes}{i1,v1}
      \fmf{dashes}{o1,v1}
      \fmf{phantom}{i2,v2,o2}
      \fmf{phantom}{i3,v3,o3}
    \end{fmfchar}
    \begin{fmfchar}(20,15)
      \fmfincoming{i1,i2,i3}
      \fmfoutgoing{o1,o2,o3}
      \fmfdot{v1,v3,v4}
      \fmf{plain,tension=0,left}{v1,v2,v1}
      \fmf{plain,tension=.2}{v3,v4}
      \fmf{dashes}{i1,v1}
      \fmf{dashes}{o1,v1}
      \fmf{phantom}{i3,v2,o3}
      \fmf{phantom}{i2,v3}
      \fmf{phantom}{o2,v4}
    \end{fmfchar}
    \begin{fmfchar}(20,20)
      \fmfincoming{i1,i2,i3,i4}
      \fmfoutgoing{o1,o2,o3,o4}
      \fmfdot{v1,v2,v3}
      \fmf{plain,tension=0,left}{v1,v2,v1}
      \fmf{plain,tension=0.5}{v2,v3}
      \fmf{plain,tension=0,left}{v3,v4,v3}
      \fmf{dashes}{i1,v1}
      \fmf{dashes}{o1,v1}
      \fmf{phantom}{i2,v2,o2}
      \fmf{phantom}{i3,v3,o3}
      \fmf{phantom}{i4,v4,o4}
    \end{fmfchar}}
\end{equation}
The sum of all snail diagrams, the ``snail'',
\begin{equation}
    \mbox{\begin{fmfchar}(20,10)
        \fmfincoming{i1,i2}
        \fmfoutgoing{o1,o2}
        \fmfblob{.5h}{v2}
        \fmfdot{v1}
        \fmf{plain,tension=0,left}{v1,v2,v1}
        \fmf{dashes}{i1,v1}
        \fmf{dashes}{o1,v1}
        \fmf{phantom}{i2,v2,o2}
      \end{fmfchar}}
\end{equation}
equals the full propagator in x-space at coinciding arguments.
In the literature the snail diagrams are sometimes also called tadpole
diagrams.
When we speak of a summation of tadpole diagrams we also include the
snail diagrams.

Tadpole and snail diagrams may appear anywhere as subdiagrams of
Feynman diagrams.
Since any Feynman diagram can be decorated by adding tadpoles and snails
at all possible places, they occur in a nested way like in this example:
\begin{equation}
    \mbox{\begin{fmfchar}(39,20)
        \fmfincoming{i1,i2,i3,i4,i5,i6}
        \fmfoutgoing{o1,o2,o3,o4,o5,o6}
        \fmfdot{v1,v2,v3,v4,v5,v6,v7,v8}
        \fmf{plain,tension=0,left}{v1,v2,h1,v2,v1}
        \fmf{dashes}{i1,v1}
        \fmf{dashes}{o1,v1}
        \fmf{phantom}{i3,v2,o3}
        \fmf{phantom}{i5,h2,h1,v6,o5}
        \fmf{plain,tension=.5,left}{v4,i3,v4}
        \fmf{plain}{v4,v3}
        \fmf{phantom,tension=.38}{v3,o2}
        \fmf{phantom}{i4,v5}
        \fmf{phantom,tension=1.6}{v5,o4}
        \fmf{plain,tension=0}{v6,v5,v7}
        \fmf{plain,tension=0,left}{v6,o6,v6}
        \fmf{phantom,tension=.5}{i3,v7}
        \fmf{plain,left}{v7,v8,o3,v8,v7}
      \end{fmfchar}}
\end{equation}
In the following we shall discuss how tadpole and snail diagrams can
be summed completely.
The discussion is in the context of bare perturbation theory.
The method, however, can be applied to renormalized perturbation theory
in the same manner.
%
\section{Dyson-Schwinger equations}
We consider the quantum field theory of a real scalar field $\phi(x)$ in
$D$-dimensional Euclidean space.
The action, containing three-point and four-point couplings, is
\begin{equation}
S = \int\!\!d^D\!x \, \left(
\frac{1}{2} ( \partial_{\mu} \phi )^2 + \frac{1}{2} m_0^2 \phi^2
+ \frac{1}{3!} f_0 \phi^3 + \frac{1}{4!} g_0 \phi^4
\right) \,.
\end{equation}
The mass and couplings are the bare ones and are therefore labelled by
an index $0$.
An action of this type appears in $\phi^4$-theory in the phase with
spontaneously broken symmetry.
In this case the relation $f_0 = \sqrt{3 g_0}\, m_0$ holds, but here we
shall not make use of it.

In coordinate space the Feynman diagrams are built from the following
elements.
The bare propagator is represented as
\begin{equation}
    \Delta(x,y) =
    \raisebox{-4\unitlength}{
      \begin{fmfchar}(10,10)
        \fmfincoming{in}
        \fmfoutgoing{out}
        \fmfdot{in,out}
        \fmf{plain}{in,out}
      \end{fmfchar}}
\nonumber
\end{equation}
and the vertices are
\begin{eqnarray}
    -g_0 \int\!\!d^D\!x & = &
    \raisebox{-4\unitlength}{
      \begin{fmfchar}(10,10)
        \fmfsurround{l1,e1,l2,e2,l3,e3,l4,e4}
        \fmfdot{v1}
        \fmf{dashes}{e1,v1}
        \fmf{dashes}{e2,v1}
        \fmf{dashes}{e3,v1}
        \fmf{dashes}{e4,v1}
      \end{fmfchar}}
       \nonumber \\
    -f_0 \int\!\!d^D\!x & = &
    \raisebox{-4\unitlength}{
      \begin{fmfchar}(10,10)
        \fmfsurround{l1,e1,l2,e2,l3,e3}
        \fmfdot{v1}
        \fmf{dashes}{e1,v1}
        \fmf{dashes}{e2,v1}
        \fmf{dashes}{e3,v1}
      \end{fmfchar}}
\nonumber
\end{eqnarray}
The connected Green's functions are denoted by
$G_c^{(n)} (x_1, x_2, \ldots, x_n)$.
Their generating functional is written as
\begin{equation}
W[j] = \ln \frac{Z[j]}{Z[0]}
\end{equation}
with
\begin{equation}
Z[j] = \int\!\!D\phi(x) \, \exp ( - S[\phi]
+ \int\!\!d^D\!x \, j(x) \phi(x) ) \,.
\end{equation}
The vertex functions $\Gamma^{(n)}(x_1, x_2, \ldots, x_n)$ are the
one-particle irreducible Green's functions with amputated external
propagators.
Their generating functional is obtained by the Legendre transformation
\begin{equation}
\Gamma[\Phi] = W[j] - \int\!\!d^D\!x \, j(x) \phi(x) \,,
\end{equation}
where
\begin{equation}
\Phi(x) := \frac{\delta W[j]}{\delta j(x)} \,, \hspace{1cm}
   j(x) = - \frac{\delta \Gamma[\Phi]}{\delta \Phi(x)} \,.
\end{equation}
The vacuum expectation value of the field $\phi(x)$ is given by
\begin{equation}
v := G_c^{(1)} (x) =
\left. \frac{\delta W[j]}{\delta j(x)} \right|_{j=0} .
\end{equation}

For the discussion of Dyson-Schwinger equations it is extremely useful
to introduce a graphical notation, see e.g.\ \cite{PC,Rivers}.
We represent the functionals graphically as follows:
\begin{eqnarray}
W[j] & \hat{=} &
    \raisebox{-3\unitlength}{
      \begin{fmfchar}(8,8)
        \fmfsurround{in,out}
        \fmfv{de.shape=circle,de.filled=.5,de.size=.8h}{v}
        \fmf{phantom}{in,v,out}
      \end{fmfchar}}
       \nonumber\\
  \Gamma[\Phi] & \hat{=} &
    \raisebox{-3\unitlength}{
      \begin{fmfchar}(8,8)
        \fmfsurround{in,out}
        \fmfv{de.shape=circle,de.filled=-.5,de.size=.8h}{v}
        \fmf{phantom}{in,v,out}
      \end{fmfchar}}
      \nonumber
  \end{eqnarray}
Their functional derivatives are associated with the following graphs:
\begin{eqnarray}
\frac{\delta W[j]}{\delta j(x_1) \, \delta j(x_2) \cdots \delta j(x_n)}
    & \hat{=} & \;\;\;
    \raisebox{-9\unitlength}{
      \begin{fmfchar*}(20,20)
        \fmfsurround{l1,en,l2,l3,e1,e2,l4,d1,d2,d3}
        \fmfv{de.shape=circle,de.filled=.5,de.size=.3h}{v}
        \fmfdot{e1,e2,en}
        \fmfv{de.shape=circle,de.filled=1,de.size=1thick}{v1,v2,v3}
        \fmf{phantom,tension=2}{d1,v1,v}
        \fmf{phantom,tension=2}{d2,v2,v}
        \fmf{phantom,tension=2}{d3,v3,v}
        \fmf{plain}{e1,v}
        \fmf{plain}{e2,v}
        \fmf{plain}{en,v}
        \fmf{phantom}{l1,v,l2}
        \fmf{phantom}{l3,v,l4}
        \fmflabel{$x_1$}{e1}
        \fmflabel{$x_2$}{e2}
        \fmflabel{$x_n$}{en}
      \end{fmfchar*}}
  \nonumber\\
  \frac{\delta \Gamma[\Phi]}
     {\delta \Phi(x_1) \, \delta \Phi(x_2) \cdots \delta \Phi(x_n)}
    & \hat{=} & \;\;\;
    \raisebox{-9\unitlength}{
      \begin{fmfchar*}(20,20)
        \fmfsurround{l1,en,l2,l3,e1,e2,l4,d1,d2,d3}
        \fmfv{de.shape=circle,de.filled=-.5,de.size=.3h}{v}
        \fmfv{de.shape=circle,de.filled=1,de.size=1thick}{v1,v2,v3}
        \fmf{phantom,tension=2}{d1,v1,v}
        \fmf{phantom,tension=2}{d2,v2,v}
        \fmf{phantom,tension=2}{d3,v3,v}
        \fmf{phantom}{l1,v,l2}
        \fmf{phantom}{l3,v,l4}
        \fmf{plain}{e1,v}
        \fmf{plain}{e2,v}
        \fmf{plain}{en,v}
        \fmflabel{$x_1$}{e1}
        \fmflabel{$x_2$}{e2}
        \fmflabel{$x_n$}{en}
      \end{fmfchar*}}
\nonumber
\end{eqnarray}
For vanishing source they are equal to the connected Green's functions
$G_c^{(n)} (x_1, x_2, \ldots, x_n)$ and the vertex functions
$\Gamma^{(n)}(x_1, x_2, \ldots, x_n)$, respectively.
The graphs for the derivatives of $\Gamma[\Phi]$ do not have dots on
their external lines, indicating that these do not represent
propagators.

For the second derivatives we have
\begin{equation}
    \raisebox{-4\unitlength}{
      \begin{fmfchar}(20,10)
        \fmfincoming{in}
        \fmfoutgoing{out}
        \fmfv{d.sh=circle,d.fi=-.5,d.si=.5h}{v1}
        \fmf{plain}{in,v1,out}
      \end{fmfchar}}
    = - \left(
      \raisebox{-4\unitlength}{
        \begin{fmfchar}(20,10)
          \fmfincoming{in}
          \fmfoutgoing{out}
          \fmfblob{.5h}{v1}
          \fmfdot{in,out}
          \fmf{plain}{in,v1,out}
        \end{fmfchar}}
    \right)^{-1}
\end{equation}

As is well known, the Feynman diagrams for
$\Gamma^{(n)}(x_1, x_2, \ldots, x_n)$
are one-particle irreducible.
It should be emphasized, however, that they may possess tadpole
subdiagrams.
These are not excluded by one-particle irreducibility.
A diagram is one-particle reducible, if by removal of an internal line
it can be separated into disconnected pieces, each of which is connected
to at least one external point.
This criterion does not apply to tadpole subdiagrams.

The Dyson-Schwinger equations \cite{Dyson,Schwinger} are the quantum
equations of motion.
They form an infinite set of coupled integro-differential equations for
the Green's functions of the theory.
It is most convenient to formulate them in terms of the generating
functionals \cite{Symanzik}, for reviews see \cite{Rivers,PC}.
{}From the identity
\begin{equation}
\int\!\!D\phi(x) \, \frac{\delta}{\delta \phi(x)}
\exp ( - S[\phi] + \int\!\!d^D\!x \, j(x) \phi(x) ) = 0
\end{equation}
one derives the Dyson-Schwinger equations for connected Green's
functions
$$
( - \partial^2 + m_0^2) \frac{\delta W}{\delta j(x)}
+ \frac{1}{2} f_0 \left[ \left( \frac{\delta W}{\delta j(x)} \right)^2
  + \frac{\delta^2 W}{\delta j(x)^2} \right]
$$
\begin{equation}
+ \frac{1}{3!} g_0 \left[
   \left( \frac{\delta W}{\delta j(x)} \right)^3
   + 3 \frac{\delta W}{\delta j(x)} \, \frac{\delta^2 W}{\delta j(x)^2}
   + \frac{\delta^3 W}{\delta j(x)^3} \right]
- j(x) = 0\,.
\end{equation}
Multiplying with a bare propagator leads to
$$
\frac{\delta W}{\delta j(x)} =
\int\!\!d^D\!x \, \Delta(x,y) \left\{
  j(y)
- \frac{1}{2} f_0 \left[ \left( \frac{\delta W}{\delta j(y)} \right)^2
  + \frac{\delta^2 W}{\delta j(y)^2} \right] \right.
$$
\begin{equation}
- \frac{1}{3!} g_0 \left. \left[
   \left( \frac{\delta W}{\delta j(y)} \right)^3
   + 3 \frac{\delta W}{\delta j(y)} \, \frac{\delta^2 W}{\delta j(y)^2}
   + \frac{\delta^3 W}{\delta j(y)^3} \right] \right\} \,.
\end{equation}
Graphically this equation reads
\begin{eqnarray}
\label{G1-Gr}
    \raisebox{-9\unitlength}{
      \begin{fmfchar}(20,20)
        \fmfincoming{in}
        \fmfoutgoing{o1}
        \fmfblob{.3h}{v1}
        \fmfdot{in}
        \fmf{plain}{in,v1}
        \fmf{phantom,tension=2}{o1,v1}
      \end{fmfchar}}
    & = &
    \raisebox{-9\unitlength}{
      \begin{fmfchar*}(15,20)
        \fmfincoming{in}
        \fmfoutgoing{o1}
        \thicklines
        \put(8,8){\line(1,1){4}}
        \put(8,12){\line(1,-1){4}}
        \fmfdot{in}
        \fmf{plain}{in,v1}
        \fmf{phantom,tension=2}{o1,v1}
      \end{fmfchar*}}
    + \frac{1}{2}
    \raisebox{-9\unitlength}{
      \begin{fmfchar}(20,20)
        \fmfincoming{in}
        \fmfoutgoing{o1,o2}
        \fmfblob{.3h}{v2,v3}
        \fmfdot{in,v1}
        \fmf{plain}{in,v1}
        \fmf{plain,tension=.5}{v3,v1,v2}
        \fmf{phantom}{o1,v2}
        \fmf{phantom}{o2,v3}
      \end{fmfchar}}
    + \frac{1}{2}
    \raisebox{-9\unitlength}{
      \begin{fmfchar}(30,20)
        \fmfincoming{in}
        \fmfoutgoing{o1}
        \fmfblob{.35h}{v2}
        \fmfdot{in,v1}
        \fmf{plain}{in,v1}
        \fmf{plain,left,tension=.4}{v2,v1,v2}
        \fmf{phantom}{o1,v2}
      \end{fmfchar}}
    \nonumber \\ & &
    + \frac{1}{3!}
    \raisebox{-9\unitlength}{
      \begin{fmfchar}(20,20)
        \fmfincoming{i1,in,i3}
        \fmfoutgoing{o1,o2,o3}
        \fmfblob{.3h}{v2,v3,v4}
        \fmfdot{in,v1}
        \fmf{plain}{in,v1,v3}
        \fmf{plain}{v2,v1,v4}
        \fmf{phantom}{i1,v2,o1}
        \fmf{phantom,tension=2}{o2,v3}
        \fmf{phantom}{i3,v4,o3}
      \end{fmfchar}}
    + \frac{1}{2}
    \raisebox{-9\unitlength}{
      \begin{fmfchar}(20,20)
        \fmfincoming{i1,in,i3}
        \fmfoutgoing{o1,o2,o3}
        \fmfblob{.3h}{v2,v3}
        \fmfdot{in,v1}
        \fmf{plain,left,tension=0.5}{v1,v2,v1}
        \fmf{plain,tension=2}{in,v1}
        \fmf{plain,tension=1}{v1,v3}
        \fmf{phantom}{i1,v2}
        \fmf{phantom,tension=2}{v2,o1}
        \fmf{phantom}{v1,o2}
        \fmf{phantom}{i3,v3}
        \fmf{phantom,tension=2}{v3,o3}
      \end{fmfchar}}
    + \frac{1}{3!}
    \raisebox{-9\unitlength}{
      \begin{fmfchar}(30,20)
        \fmfincoming{in}
        \fmfoutgoing{o1}
        \fmfblob{.3w}{v2}
        \fmfdot{v1}
        \fmfdot{in}
        \fmf{plain,tension=1.5}{in,v1}
        \fmf{plain,left,tension=0}{v2,v1,v2}
        \fmf{plain,straight}{v1,v2}
        \fmf{phantom}{o1,v2}
      \end{fmfchar}}
\end{eqnarray}
where the source is represented by a cross:
$$
    \int\!\!d^D\!x \, j(x) =
    \raisebox{-4\unitlength}{
      \begin{picture}(10,10)
        \thicklines
        \put(3,3){\line(1,1){4}}
        \put(3,7){\line(1,-1){4}}
      \end{picture}}
$$
The Dyson-Schwinger equations for Green's functions are now obtained by
differentiating this equation several times with respect to the source
and setting $j=0$ afterwards.
Differentiating once yields
\begin{eqnarray}
\label{G2-Gr}
    \raisebox{-9\unitlength}{
      \begin{fmfchar}(20,20)
        \fmfincoming{in}
        \fmfoutgoing{out}
        \fmfblob{.3h}{v1}
        \fmfdot{in,out}
        \fmf{plain}{in,v1,out}
      \end{fmfchar}}
    & = &
    \raisebox{-9\unitlength}{
      \begin{fmfchar}(10,20)
        \fmfincoming{in}
        \fmfoutgoing{out}
        \fmfdot{in,out}
        \fmf{plain}{in,out}
      \end{fmfchar}}
    \: +
    \raisebox{-9\unitlength}{
      \begin{fmfchar}(30,20)
        \fmfincoming{i1,in,i2}
        \fmfoutgoing{o1,out,o2}
        \fmfblob{.3h}{v2,v3}
        \fmfdot{in,v1,out}
        \fmf{plain}{in,v1,v2,out}
        \fmf{plain,tension=0}{v1,v3}
        \fmf{phantom}{i2,v3,v4,o2}
      \end{fmfchar}}
    \: + \frac{1}{2}
    \raisebox{-9\unitlength}{
      \begin{fmfchar}(30,20)
        \fmfincoming{i1,in,i2}
        \fmfoutgoing{o1,out,o2}
        \fmfblob{.3h}{v2,v3,v4}
        \fmfdot{in,v1,out}
        \fmf{plain}{in,v1,v2,out}
        \fmf{plain,tension=0}{v3,v1,v4}
        \fmf{phantom}{i2,v3,v5,v6,v4,v7,v8,o2}
      \end{fmfchar}}
    \nonumber \\ & &
    + \frac{1}{2}
    \raisebox{-9\unitlength}{
      \begin{fmfchar}(30,20)
        \fmfincoming{i1,in,i2}
        \fmfoutgoing{o1,out,o2}
        \fmfblob{.3h}{v2,v3}
        \fmfdot{in,v1,out}
        \fmf{plain}{in,v1,v2,out}
        \fmf{plain,left,tension=0}{v1,v3,v1}
        \fmf{phantom}{i2,v3,v4,o2}
      \end{fmfchar}}
    \: + \frac{1}{2}
    \raisebox{-9\unitlength}{
      \begin{fmfchar}(30,20)
        \fmfincoming{in}
        \fmfoutgoing{out}
        \fmfblob{.3h}{v2}
        \fmfdot{in,v1,out}
        \fmf{plain}{in,v1}
        \fmf{plain}{v2,out}
        \fmf{plain,left,tension=.5}{v1,v2,v1}
      \end{fmfchar}}
    \nonumber \\ & &
    + \frac{1}{2}
    \raisebox{-9\unitlength}{
      \begin{fmfchar}(30,20)
        \fmfincoming{i1,in,i2}
        \fmfoutgoing{o1,out,o2}
        \fmfblob{.3h}{v2,v3}
        \fmfdot{in,v1,out}
        \fmf{plain}{in,v1}
        \fmf{plain,tension=0}{v1,v3}
        \fmf{plain}{v2,out}
        \fmf{plain,left,tension=.5}{v1,v2,v1}
        \fmf{phantom}{i2,v3,v4,v5,o2}
      \end{fmfchar}}
    \: + \frac{1}{3!}
    \raisebox{-9\unitlength}{
      \begin{fmfchar}(30,20)
        \fmfincoming{in}
        \fmfoutgoing{out}
        \fmfblob{.3h}{v2}
        \fmfdot{in,v1,out}
        \fmf{plain}{in,v1,v2,out}
        \fmf{plain,left,tension=0}{v1,v2,v1}
      \end{fmfchar}}
\end{eqnarray}
For $j = 0$ this is an equation for the connected two-point function,
i.e.\ the full propagator, but it also holds for nonvanishing source.
Further differentiation yields the hierarchy of Dyson-Schwinger
equations.

The perturbation series in terms of Feynman diagrams can be obtained by
solving the Dyson-Schwinger equations iteratively, see e.g.\
\cite{Rivers}.
For example, the perturbative expansion of the full propagator is
obtained from (\ref{G2-Gr}) by replacing the occurring Green's functions
on the right hand side by the expressions (\ref{G1-Gr},\ref{G2-Gr}) and
the corresponding higher ones, and repeating this substitution
successively.
{}From Eqs.\ (\ref{G1-Gr},\ref{G2-Gr}) one sees that all sorts of nested
tadpole and snail diagrams will be generated in this way.

We shall also need the Dyson-Schwinger equations for the vertex
functions.
In functional form they are
$$
- \frac{\delta \Gamma[\Phi]}{\delta \Phi(x)} =
( - \partial^2 + m_0^2 ) \Phi(x)
+ \frac{1}{2} f_0
   \left[ \Phi^2(x) + \frac{\delta^2 W}{\delta j(x) \delta j(x)} \right]
$$
$$
+ \frac{1}{3!} g_0 \left[ \Phi^3(x)
          + 3 \Phi(x) \frac{\delta^2 W}{\delta j(x) \delta j(x)} \right.
$$
\begin{equation}
+ \int\!\!d^D\!z_1 d^D\!z_2 d^D\!z_3 \left.
       \frac{\delta^2 W}{\delta j(x) \delta j(z_1)} \,
       \frac{\delta^2 W}{\delta j(x) \delta j(z_2)} \,
       \frac{\delta^2 W}{\delta j(x) \delta j(z_3)} \,
       \frac{\delta^3 \Gamma}
                  {\delta \Phi(z_1) \delta \Phi(z_2) \delta \Phi(z_3)}
          \right] ,
\end{equation}
or graphically
\begin{eqnarray}
    \raisebox{-9\unitlength}{
      \begin{fmfchar}(20,20)
        \fmfincoming{in}
        \fmfoutgoing{o1}
        \fmfv{d.sh=circle,d.fi=-.5,d.si=.3h}{v1}
        \fmf{plain}{in,v1}
        \fmf{phantom,tension=2}{o1,v1}
      \end{fmfchar}}
    & = & -
    \raisebox{-9\unitlength}{
      \begin{fmfchar*}(20,20)
        \fmfincoming{in}
        \fmfoutgoing{o1}
        \fmfblob{.3h}{v1}
        \fmf{plain}{in,v1}
        \fmf{phantom,tension=2}{o1,v1}
        \thicklines
        \put(4,7){\line(1,2){3}}
      \end{fmfchar*}}
    + \frac{1}{2}
    \raisebox{-9\unitlength}{
      \begin{fmfchar}(20,20)
        \fmfincoming{in}
        \fmfoutgoing{o1,o2}
        \fmfblob{.3h}{v2,v3}
        \fmfdot{v1}
        \fmf{dashes}{in,v1}
        \fmf{plain,tension=.5}{v3,v1,v2}
        \fmf{phantom}{o1,v2}
        \fmf{phantom}{o2,v3}
      \end{fmfchar}}
    + \frac{1}{2}
    \raisebox{-9\unitlength}{
      \begin{fmfchar}(30,20)
        \fmfincoming{in}
        \fmfoutgoing{o1}
        \fmfblob{.35h}{v2}
        \fmfdot{v1}
        \fmf{dashes}{in,v1}
        \fmf{plain,left,tension=.4}{v2,v1,v2}
        \fmf{phantom}{o1,v2}
      \end{fmfchar}}
    \nonumber \\ & &
    + \frac{1}{3!}
    \raisebox{-9\unitlength}{
      \begin{fmfchar}(20,20)
        \fmfincoming{i1,in,i3}
        \fmfoutgoing{o1,o2,o3}
        \fmfblob{.3h}{v2,v3,v4}
        \fmfdot{v1}
        \fmf{dashes}{in,v1}
        \fmf{plain}{v1,v3}
        \fmf{plain}{v2,v1,v4}
        \fmf{phantom}{i1,v2,o1}
        \fmf{phantom,tension=2}{o2,v3}
        \fmf{phantom}{i3,v4,o3}
      \end{fmfchar}}
    + \frac{1}{2}
    \raisebox{-9\unitlength}{
      \begin{fmfchar}(20,20)
        \fmfincoming{i1,in,i3}
        \fmfoutgoing{o1,o2,o3}
        \fmfblob{.3h}{v2,v3}
        \fmfdot{v1}
        \fmf{plain,left,tension=0.5}{v1,v2,v1}
        \fmf{dashes,tension=2}{in,v1}
        \fmf{plain,tension=1}{v1,v3}
        \fmf{phantom}{i1,v2}
        \fmf{phantom,tension=2}{v2,o1}
        \fmf{phantom}{v1,o2}
        \fmf{phantom}{i3,v3}
        \fmf{phantom,tension=2}{v3,o3}
      \end{fmfchar}}
    + \frac{1}{3!}
    \raisebox{-9\unitlength}{
      \begin{fmfchar}(30,20)
        \fmfincoming{i1,i2,i3}
        \fmfoutgoing{o1,o2,o3}
        \fmfblob{.25h}{v2,v3,v4}
        \fmfv{d.sh=circle,d.fi=-.5,d.si=.4h}{v5}
        \fmfdot{v1}
        \fmf{dashes,tension=2}{i2,v1}
        \fmf{plain,tension=0.2}{v1,v3,v5,v4,v1}
        \fmf{plain}{v1,v2,v5}
        \fmf{phantom,tension=2}{o2,v5}
        \fmf{phantom}{i1,v3,o1}
        \fmf{phantom}{i3,v4,o3}
      \end{fmfchar}}
\end{eqnarray}
Here the inverse bare propagator is denoted by a stroked line:
\begin{equation}
\Delta^{-1}(x,y) = (-\partial^2+m_0^2)\delta(x-y) =
    \raisebox{-9\unitlength}{
      \begin{fmfchar*}(10,20)
        \fmfincoming{in}
        \fmfoutgoing{out}
        \fmf{plain}{in,out}
        \thicklines
        \put(3.5,7){\line(1,2){3}}
      \end{fmfchar*}} \quad .
\end{equation}
Again, differentiating several times with respect to $\Phi(x)$ yields a
hierarchy of equations for the vertex functions.
For this purpose one has to use
\begin{equation}
\label{diffg2}
\frac{\delta}{\delta \Phi(x)}\raisebox{-9\unitlength}{
      \begin{fmfchar}(20,20)
        \fmfincoming{in}
        \fmfoutgoing{out}
        \fmfblob{.25h}{v1}
        \fmfdot{in,out}
        \fmf{plain}{in,v1,out}
      \end{fmfchar}}
     = \raisebox{-9\unitlength}{
      \begin{fmfchar*}(39,20)
        \fmfsurround{out,l,in,e}
        \fmfblob{.25h}{v1,v3}
        \fmfv{d.sh=circle,d.fi=-.5,d.si=.25h}{v2}
        \fmfdot{in,out}
        \fmf{plain}{in,v1,v2,v3,out}
        \fmf{plain,tension=0}{e,v2}
        \fmflabel{$x$}{e}
      \end{fmfchar*}}
\end{equation}
For the first functional derivative we obtain
\begin{eqnarray}
\label{OPI-2PktGr}
    \raisebox{-9\unitlength}{
      \begin{fmfchar}(20,20)
        \fmfincoming{in}
        \fmfoutgoing{out}
        \fmfv{d.sh=circle,d.fi=-.5,d.si=.3h}{v1}
        \fmf{plain}{in,v1,out}
      \end{fmfchar}}
    & = & -
    \raisebox{-9\unitlength}{
      \begin{fmfchar*}(10,20)
        \fmfincoming{in}
        \fmfoutgoing{out}
        \fmf{plain}{in,out}
        \thicklines
        \put(3.5,7){\line(1,2){3}}
      \end{fmfchar*}}
    \: +
    \raisebox{-9\unitlength}{
      \begin{fmfchar}(20,20)
        \fmfincoming{i1,in,i2}
        \fmfoutgoing{o1,out,o2}
        \fmfblob{.3h}{v2}
        \fmfdot{v1}
        \fmf{plain,tension=0}{v1,v2}
        \fmf{dashes}{in,v1}
        \fmf{dashes}{out,v1}
        \fmf{phantom}{i2,v2,o2}
      \end{fmfchar}}
    \: + \frac{1}{2}
    \raisebox{-9\unitlength}{
      \begin{fmfchar}(20,20)
        \fmfincoming{i1,in,i2}
        \fmfoutgoing{o1,out,o2}
        \fmfblob{.3h}{v2,v3}
        \fmfdot{v1}
        \fmf{plain,tension=0}{v2,v1,v3}
        \fmf{dashes}{in,v1}
        \fmf{dashes}{out,v1}
        \fmf{phantom}{i2,v2,v4,v3,o2}
      \end{fmfchar}}
    \: + \frac{1}{2}
    \raisebox{-9\unitlength}{
      \begin{fmfchar}(20,20)
        \fmfincoming{i1,in,i2}
        \fmfoutgoing{o1,out,o2}
        \fmfblob{.3h}{v2}
        \fmfdot{v1}
        \fmf{plain,tension=0,left}{v1,v2,v1}
        \fmf{dashes}{in,v1}
        \fmf{dashes}{out,v1}
        \fmf{phantom}{i2,v2,o2}
      \end{fmfchar}}
    \nonumber \\ & &
    + \frac{1}{2}
    \raisebox{-9\unitlength}{
      \begin{fmfchar}(30,20)
        \fmfincoming{i1,in,i2}
        \fmfoutgoing{o1,out,o2}
        \fmfblob{.25h}{v2,v3}
        \fmfv{d.sh=circle,d.fi=-.5,d.si=.3h}{v4}
        \fmfdot{v1}
        \fmf{dashes}{in,v1}
        \fmf{plain,tension=.5}{v4,v2,v1,v3,v4}
        \fmf{plain}{v4,out}
        \fmf{phantom}{i1,v2,o1}
        \fmf{phantom}{i2,v3,o2}
      \end{fmfchar}}
    \: + \frac{1}{2}
    \raisebox{-9\unitlength}{
      \begin{fmfchar}(30,20)
        \fmfincoming{i1,in,i2}
        \fmfoutgoing{o1,out,o2}
        \fmfblob{.25h}{v2,v3,v4}
        \fmfv{d.sh=circle,d.fi=-.5,d.si=.3h}{v5}
        \fmfdot{v1}
        \fmf{dashes}{in,v1}
        \fmf{plain,tension=.5}{v5,v2,v1,v3,v5}
        \fmf{plain,tension=0}{v1,v4}
        \fmf{plain}{v5,out}
        \fmf{phantom}{i1,v2,o1}
        \fmf{phantom}{i2,v3,o2}
        \fmf{phantom,tension=6}{i2,v4}
        \fmf{phantom}{v4,o2}
      \end{fmfchar}}
    \nonumber \\ & &
    + \frac{1}{2}
    \raisebox{-9\unitlength}{
      \begin{fmfchar}(30,20)
        \fmfsurround{out,a3,a2,a1,in,a6,a5,a4}
        \fmfblob{.2h}{v4,v5,v6,v7}
        \fmfv{d.sh=circle,d.fi=-.5,d.si=.25h}{v2,v3}
        \fmfdot{v1}
        \fmf{dashes}{in,v1}
        \fmf{plain,tension=.4}{v1,v4,v3}
        \fmf{phantom,tension=.4}{v1,v2,v3}
        \fmf{plain,tension=0}{v2,v6,v1,v7,v2,v5,v3}
        \fmf{plain}{v3,out}
        \fmf{phantom,tension=2}{a5,v2}
        \fmf{phantom,tension=2}{a2,v4}
        \fmf{phantom}{a4,v5,va,vb,vc,in}
        \fmf{phantom}{a6,v6,v7,v8,v9,a3}
      \end{fmfchar}}
    \: + \frac{1}{3!}
    \raisebox{-9\unitlength}{
      \begin{fmfchar}(30,20)
        \fmfincoming{i1,i2,i3}
        \fmfoutgoing{o1,o2,o3}
        \fmfblob{.25h}{v2,v3,v4}
        \fmfv{d.sh=circle,d.fi=-.5,d.si=.3h}{v5}
        \fmfdot{v1}
        \fmf{dashes,tension=2}{i2,v1}
        \fmf{plain,tension=0.2}{v1,v3,v5,v4,v1}
        \fmf{plain}{v1,v2,v5}
        \fmf{plain,tension=2}{v5,o2}
        \fmf{phantom}{i1,v3,o1}
        \fmf{phantom}{i3,v4,o3}
      \end{fmfchar}}
\end{eqnarray}
This is the Dyson-Schwinger equation for the inverse propagator.
%
\section{Elimination of tadpoles from vertex functions}
Equation (\ref{OPI-2PktGr}) for the inverse propagator contains the
tadpole, the double-tadpole and the snail as constant terms on the right
hand side.
The first term on the right hand side is the inverse bare propagator
$- ( - \partial^2 + m_0^2) \delta( x-y )$.
This suggests to introduce the modified mass
\begin{equation}
\tilde{m}_0^2 := m_0^2 - A(m_0,f_0,g_0) \,,
\end{equation}
where
\begin{equation}
A := A_1 + A_2 + A_3
\end{equation}
\begin{eqnarray}
\label{AdefGr}
A_1 & := &
    \raisebox{-9\unitlength}{
      \begin{fmfchar}(20,20)
        \fmfincoming{i1,in,i2}
        \fmfoutgoing{o1,out,o2}
        \fmfblob{.3h}{v2}
        \fmfdot{v1}
        \fmf{plain,tension=0}{v1,v2}
        \fmf{dashes}{in,v1}
        \fmf{dashes}{out,v1}
        \fmf{phantom}{i2,v2,o2}
      \end{fmfchar}}
    = -f_0 \frac{\delta W}{\delta j(x)} \nonumber \\
  A_2 & := & \frac{1}{2}
    \raisebox{-9\unitlength}{
      \begin{fmfchar}(20,20)
        \fmfincoming{i1,in,i2}
        \fmfoutgoing{o1,out,o2}
        \fmfblob{.3h}{v2,v3}
        \fmfdot{v1}
        \fmf{plain,tension=0}{v2,v1,v3}
        \fmf{dashes}{in,v1}
        \fmf{dashes}{out,v1}
        \fmf{phantom}{i2,v2,v4,v3,o2}
      \end{fmfchar}}
    = -\frac{1}{2} g_0 \left( \frac{\delta W}{\delta j(x)} \right)^2 \\
  A_3 & := & \frac{1}{2}
    \raisebox{-9\unitlength}{
      \begin{fmfchar}(20,20)
        \fmfincoming{i1,in,i2}
        \fmfoutgoing{o1,out,o2}
        \fmfblob{.3h}{v2}
        \fmfdot{v1}
        \fmf{plain,tension=0,left}{v1,v2,v1}
        \fmf{dashes}{in,v1}
        \fmf{dashes}{out,v1}
        \fmf{phantom}{i2,v2,o2}
      \end{fmfchar}}
    = - \frac{1}{2} g_0 \frac{\delta^2 W}{\delta j(x)^2} \nonumber \,,
\end{eqnarray}
as well as the modified propagator
\begin{equation}
\tilde\Delta(x,y) := (- \partial^2+\tilde{m}_0^2)^{-1}(x,y) =
    \raisebox{-4\unitlength}{
      \begin{fmfchar}(10,10)
        \fmfincoming{in}
        \fmfoutgoing{out}
        \fmfdot{in,out}
        \fmf{photon}{in,out}
      \end{fmfchar}}
\end{equation}
For vanishing source we have
\begin{equation}
A_1 = - f_0 v \,, \hspace{1cm}
A_2 = - \frac{1}{2} g_0 v^2 \,, \hspace{1cm}
A_3 = - \frac{1}{2} g_0 G_c^{(2)} (x,x) \,.
\end{equation}
The sixth term on the right hand side of (\ref{OPI-2PktGr}) contains
another tadpole attached to a bare vertex.
We take this one into account by defining a modified bare three-point
vertex $\tilde{f_0}$ by means of
\begin{equation}
\label{VertF2}
-\tilde{f_0} =
    \raisebox{-4\unitlength}{
      \begin{fmfchar}(10,10)
        \fmfsurround{l1,e1,l2,e2,l3,e3}
        \fmftil{v1}
        \fmf{dashes}{e1,v1}
        \fmf{dashes}{e2,v1}
        \fmf{dashes}{e3,v1}
      \end{fmfchar}}
  :=
    \raisebox{-4\unitlength}{
      \begin{fmfchar}(10,10)
        \fmfsurround{l1,e1,l2,e2,l3,e3}
        \fmfdot{v1}
        \fmf{dashes}{e1,v1}
        \fmf{dashes}{e2,v1}
        \fmf{dashes}{e3,v1}
      \end{fmfchar}}
    \: +
    \raisebox{-4\unitlength}{
      \begin{fmfchar}(10,10)
        \fmfsurround{l1,e1,l2,e2,l3,e3}
        \fmfdot{v1}
        \fmfblob{.4h}{l2}
        \fmf{dashes}{e1,v1}
        \fmf{dashes}{e2,v1}
        \fmf{dashes}{e3,v1}
        \fmf{plain,tension=0}{l2,v1}
      \end{fmfchar}}
    \: = - f_0 - g_0 \frac{\delta W}{\delta j(x)} \,.
\end{equation}
With these definitions the Dyson-Schwinger equation for the two-point
vertex function can be written
\begin{eqnarray}
\label{mod-OPI-2PktGr}
    \raisebox{-9\unitlength}{
      \begin{fmfchar}(20,20)
        \fmfincoming{in}
        \fmfoutgoing{out}
        \fmfv{d.sh=circle,d.fi=-.5,d.si=.3h}{v1}
        \fmf{plain}{in,v1,out}
      \end{fmfchar}}
    & = & - (
    \raisebox{-9\unitlength}{
      \begin{fmfchar}(10,20)
        \fmfincoming{in}
        \fmfoutgoing{out}
        \fmfdot{in,out}
        \fmf{photon}{in,out}
      \end{fmfchar}}
    \: )^{-1}
    \: + \frac{1}{2}
    \raisebox{-9\unitlength}{
      \begin{fmfchar}(30,20)
        \fmfincoming{i1,in,i2}
        \fmfoutgoing{o1,out,o2}
        \fmfblob{.25h}{v2,v3}
        \fmfv{d.sh=circle,d.fi=-.5,d.si=.3h}{v4}
        \fmftil{v1}
        \fmf{dashes}{in,v1}
        \fmf{plain,tension=.5}{v4,v2,v1,v3,v4}
        \fmf{plain}{v4,out}
        \fmf{phantom}{i1,v2,o1}
        \fmf{phantom}{i2,v3,o2}
      \end{fmfchar}}
    \nonumber \\ & &
    + \frac{1}{2}
    \raisebox{-9\unitlength}{
      \begin{fmfchar}(30,20)
        \fmfsurround{out,a3,a2,a1,in,a6,a5,a4}
        \fmfblob{.2h}{v4,v5,v6,v7}
        \fmfv{d.sh=circle,d.fi=-.5,d.si=.25h}{v2,v3}
        \fmfdot{v1}
        \fmf{dashes}{in,v1}
        \fmf{plain,tension=.4}{v1,v4,v3}
        \fmf{phantom,tension=.4}{v1,v2,v3}
        \fmf{plain,tension=0}{v2,v6,v1,v7,v2,v5,v3}
        \fmf{plain}{v3,out}
        \fmf{phantom,tension=2}{a5,v2}
        \fmf{phantom,tension=2}{a2,v4}
        \fmf{phantom}{a4,v5,va,vb,vc,in}
        \fmf{phantom}{a6,v6,v7,v8,v9,a3}
      \end{fmfchar}}
    \: + \frac{1}{3!}
    \raisebox{-9\unitlength}{
      \begin{fmfchar}(30,20)
        \fmfincoming{i1,i2,i3}
        \fmfoutgoing{o1,o2,o3}
        \fmfblob{.25h}{v2,v3,v4}
        \fmfv{d.sh=circle,d.fi=-.5,d.si=.3h}{v5}
        \fmfdot{v1}
        \fmf{dashes,tension=2}{i2,v1}
        \fmf{plain,tension=0.2}{v1,v3,v5,v4,v1}
        \fmf{plain}{v1,v2,v5}
        \fmf{plain,tension=2}{v5,o2}
        \fmf{phantom}{i1,v3,o1}
        \fmf{phantom}{i3,v4,o3}
      \end{fmfchar}}
\end{eqnarray}
Equation (\ref{G2-Gr}) for the full propagator correspondingly goes over
into
\begin{equation}
\label{mod-G2-Gr}
    \raisebox{-9\unitlength}{
      \begin{fmfchar}(20,20)
        \fmfincoming{in}
        \fmfoutgoing{out}
        \fmfblob{.3h}{v1}
        \fmfdot{in,out}
        \fmf{plain}{in,v1,out}
      \end{fmfchar}}
    =
    \raisebox{-9\unitlength}{
      \begin{fmfchar}(10,20)
        \fmfincoming{in}
        \fmfoutgoing{out}
        \fmfdot{in,out}
        \fmf{photon}{in,out}
      \end{fmfchar}}
    \: + \frac{1}{2}
    \raisebox{-9\unitlength}{
      \begin{fmfchar}(30,20)
        \fmfincoming{in}
        \fmfoutgoing{out}
        \fmfblob{.3h}{v2}
        \fmfdot{in,out}
        \fmftil{v1}
        \fmf{photon}{in,v1}
        \fmf{plain}{v2,out}
        \fmf{plain,left,tension=.5}{v1,v2,v1}
      \end{fmfchar}}
    \: + \frac{1}{3!}
    \raisebox{-9\unitlength}{
      \begin{fmfchar}(30,20)
        \fmfincoming{in}
        \fmfoutgoing{out}
        \fmfblob{.3h}{v2}
        \fmfdot{in,v1,out}
        \fmf{photon}{in,v1}
        \fmf{plain}{v1,v2,out}
        \fmf{plain,left,tension=0}{v1,v2,v1}
      \end{fmfchar}}
\end{equation}
The modified Dyson-Schwinger equations for $\Gamma^{(2)}$ and
$G_c^{(2)}$ do not contain any tadpoles or snails.
The next step is to show that the same holds for all higher vertex
functions $\Gamma^{(n)}$ with $n \geq 2$.
This can be shown by induction on $n$.
Assume that it is true for $\Gamma^{(k)}$ with $k < n$.
The equation for $\Gamma^{(n)}$ is obtained from the one for
$\Gamma^{(n-1)}$ (for nonvanishing source $j(x)$) by differentiation
with respect to $\Phi(x)$:
\begin{equation}
\label{diffGamma}
\frac{\delta}{\delta \Phi(x)} \hspace{5mm}
        \raisebox{-9\unitlength}{
          \begin{fmfchar}(20,20)
            \fmfincoming{i6,i5,i4,i3,i2,i1}
            \fmfoutgoing{out}
            \fmfgam{.4h}{v1}
            \fmfdot{i1,i2,i6}
            \fmfv{de.shape=circle,de.filled=1,de.size=1thick}{v3,v4,v5}
            \fmf{plain}{i1,v1,i2}
            \fmf{plain}{i6,v1}
            \fmf{phantom,tension=.5}{i3,v3,v1,v4,i4}
            \fmf{phantom,tension=.5}{i5,v5,v1}
            \fmf{phantom,tension=6}{v1,out}
          \end{fmfchar}}
      =
        \raisebox{-9\unitlength}{
          \begin{fmfchar*}(20,20)
            \fmfincoming{i6,i5,i4,i3,i2,i1}
            \fmfoutgoing{out}
            \fmfgam{.4h}{v1}
            \fmfdot{i1,i2,i6,out}
            \fmfv{de.shape=circle,de.filled=1,de.size=1thick}{v3,v4,v5}
            \fmf{plain}{i1,v1,i2}
            \fmf{plain}{i6,v1}
            \fmf{phantom,tension=.5}{i3,v3,v1,v4,i4}
            \fmf{phantom,tension=.5}{i5,v5,v1}
            \fmf{plain,tension=6}{v1,out}
            \put(22,10){\makebox(0,0)[l]{$x$}}
          \end{fmfchar*}}
\end{equation}
The Dyson-Schwinger equation for $\Gamma^{(n-1)}$ contains propagators
$G_c^{(2)}$ and lower vertex functions $\Gamma^{(k)}$, $k < n-1$ on its
right hand side.
Differentiating any propagator does not produce tadpoles or snails, as
follows from Eq.\ (\ref{diffg2}).
Differentiating any $\Gamma^{(k)}$, $k < n-1$, also does not produce
them according to (\ref{diffGamma}).
{}From the fact that the proposition is true for $n = 2$ its validity
for all $n$ follows.

Now we make use of the fact that iteration of the Dyson-Schwinger
equations produces perturbation theory in terms of Feynman diagrams, as
discussed above.
Using the modified form of the equations we conclude that it is possible
to calculate $\Gamma^{(n)}$, $n \geq 2$, in terms of Feynman diagrams
without tadpoles or snails.
The procedure is summarized as follows:
\begin{itemize}
\item
In order to calculate $\Gamma^{(n)}$ draw all one-particle irreducible
Feynman diagrams which do not contain tadpole or snail subdiagrams.
\item
For the lines use the modified propagator
$\tilde\Delta = (- \partial^2+\tilde{m}_0^2)^{-1}$
with $\tilde{m}_0^2 := m_0^2 - A$.
\item
The four-point vertex is given by $-g_0$,\\
the three-point vertex by $-\tilde{f_0} = - f_0 - g_0 v$.
\end{itemize}
The tadpoles have been summed up effectively by the modification of the
bare parameters.
What remains to be done is the calculation of $A$ and $v$ in terms of
Feynman diagrams without tadpoles.
%
\section{Calculation of tadpoles}
In order to express the vertex functions $\Gamma^{(n)}$, $n \geq 2$, in
terms of the original bare parameters $m_0$ and $g_0$, which is often
necessary in applications (see e.g.\ \cite{GKM}), we have to substitute
\begin{eqnarray}
\label{tilde-def}
\tilde{m}_0^2 & = & m_0^2 + f_0 v + \frac{1}{2} g_0 v^2
   + \frac{1}{2} g_0 G_c^{(2)}(x,x) \nonumber\\
\tilde{f_0} & = & f_0 + g_0 v \,.
\end{eqnarray}
The two-point function $G_c^{(2)}(x,x)$ can be obtained as described in
the previous section.
So the tadpole $v$ itself must finally be calculated.
It is given by the sum of all connected Feynman diagrams with one
external point.
It suggests itself to apply the summation method again and to calculate
$v$ by excluding tadpole subdiagrams and using the modified parameters
$\tilde{m}_0$ and $\tilde{f_0}$.
This procedure, however, does not yield the correct result.
The combinatoric factors would come out wrong at two loops and beyond,
as an explicit calculation shows.
The reason is that the symmetry factors in the appropriate
Dyson-Schwinger equation for the one-point function are not those which
would correspond to a naive application of the modified Feynman rules.
Let us consider this equation:
\begin{eqnarray}
    \raisebox{-9\unitlength}{
      \begin{fmfchar}(20,20)
        \fmfincoming{in}
        \fmfoutgoing{o1}
        \fmfblob{.3h}{v1}
        \fmfdot{in}
        \fmf{plain}{in,v1}
        \fmf{phantom,tension=2}{o1,v1}
      \end{fmfchar}}
    & = &
      \frac{1}{2}
    \raisebox{-9\unitlength}{
      \begin{fmfchar}(30,20)
        \fmfincoming{in}
        \fmfoutgoing{o1}
        \fmfblob{.35h}{v2}
        \fmfdot{in,v1}
        \fmf{plain}{in,v1}
        \fmf{plain,left,tension=.4}{v2,v1,v2}
        \fmf{phantom}{o1,v2}
      \end{fmfchar}}
    + \frac{1}{3!}
    \raisebox{-9\unitlength}{
      \begin{fmfchar}(30,20)
        \fmfincoming{in}
        \fmfoutgoing{o1}
        \fmfblob{.3w}{v2}
        \fmfdot{v1}
        \fmfdot{in}
        \fmf{plain,tension=1.5}{in,v1}
        \fmf{plain,left,tension=0}{v2,v1,v2}
        \fmf{plain,straight}{v1,v2}
        \fmf{phantom}{o1,v2}
      \end{fmfchar}}
    \nonumber \\ & &
    + \frac{1}{2}
    \raisebox{-9\unitlength}{
      \begin{fmfchar}(20,20)
        \fmfincoming{i1,in,i3}
        \fmfoutgoing{o1,o2,o3}
        \fmfblob{.3h}{v2,v3}
        \fmfdot{in,v1}
        \fmf{plain,left,tension=0.5}{v1,v2,v1}
        \fmf{plain,tension=2}{in,v1}
        \fmf{plain,tension=1}{v1,v3}
        \fmf{phantom}{i1,v2}
        \fmf{phantom,tension=2}{v2,o1}
        \fmf{phantom}{v1,o2}
        \fmf{phantom}{i3,v3}
        \fmf{phantom,tension=2}{v3,o3}
      \end{fmfchar}}
    + \frac{1}{2}
    \raisebox{-9\unitlength}{
      \begin{fmfchar}(20,20)
        \fmfincoming{in}
        \fmfoutgoing{o1,o2}
        \fmfblob{.3h}{v2,v3}
        \fmfdot{in,v1}
        \fmf{plain}{in,v1}
        \fmf{plain,tension=.5}{v3,v1,v2}
        \fmf{phantom}{o1,v2}
        \fmf{phantom}{o2,v3}
      \end{fmfchar}}
    + \frac{1}{3!}
    \raisebox{-9\unitlength}{
      \begin{fmfchar}(20,20)
        \fmfincoming{i1,in,i3}
        \fmfoutgoing{o1,o2,o3}
        \fmfblob{.3h}{v2,v3,v4}
        \fmfdot{in,v1}
        \fmf{plain}{in,v1,v3}
        \fmf{plain}{v2,v1,v4}
        \fmf{phantom}{i1,v2,o1}
        \fmf{phantom,tension=2}{o2,v3}
        \fmf{phantom}{i3,v4,o3}
      \end{fmfchar}}
\end{eqnarray}
\begin{equation}
\label{DS-TP}
v = - \frac{f_0}{2 m_0^2} G_c^{(2)}
    - \frac{g_0}{6 m_0^2} G_c^{(3)}
    - \frac{g_0}{2 m_0^2} G_c^{(2)} v
    - \frac{f_0}{2 m_0^2} v^2
    - \frac{g_0}{6 m_0^2} v^3 \,,
\end{equation}
with the abbreviations
\begin{equation}
G_c^{(2)} := G_c^{(2)}(x,x) \,, \hspace{1cm}
G_c^{(3)} := G_c^{(2)}(x,x,x) \,.
\end{equation}
This is a cubic equation for $v$, of the form
\begin{equation}
v = a + b v^2 + c v^3
\end{equation}
with
\begin{equation}
b = - \left( 1 + \frac{g_0}{2 m_0^2} G_c^{(2)} \right)^{-1}
    \frac{f_0}{2 m_0^2} \,, \hspace{5mm}
c = - \left( 1 + \frac{g_0}{2 m_0^2} G_c^{(2)} \right)^{-1}
    \frac{g_0}{6 m_0^2} \,, \hspace{5mm}
a = b G_c^{(2)} + c G_c^{(3)} \,.
\end{equation}
The relevant solution can be written down in closed form with the help
of Cardano's formulae \cite{Cardano} but for the purpose of perturbation
theory it is more convenient to write it as a power series in the
couplings:
\begin{eqnarray}
v & = &
  - \frac{f_0}{2 m_0^2} G_c^{(2)}
  - \frac{g_0}{6 m_0^2} G_c^{(3)}
  + \frac{f_0}{2 m_0^2} \frac{g_0}{2 m_0^2} ( G_c^{(2)} )^2
  + \frac{1}{3} \left( \frac{g_0}{2 m_0^2} \right)^2 G_c^{(2)} G_c^{(3)}
  \nonumber\\
  &   &
  - \frac{f_0}{2 m_0^2} \left( \frac{g_0}{2 m_0^2} \right)^2
    ( G_c^{(2)} )^3
  + \ldots \\
  & =: & F( G_c^{(2)}, G_c^{(3)}, m_0, g_0, f_0 ) \,. \nonumber
\end{eqnarray}
This series is most easily obtained by solving (\ref{DS-TP})
iteratively.

Now $v$ is known in terms of $G_c^{(2)}$ and $G_c^{(3)}$, which on the
other hand are calculable by means of the modified perturbation theory.
In practice one proceeds in the following way:
\begin{itemize}
\item
Calculate $G_c^{(2)}$ and $G_c^{(3)}$ in terms of $\tilde{m}_0$, $g_0$
and $\tilde{f_0}$ with the modified Feynman rules of the previous
section.
\item
Insert the definitions (\ref{tilde-def}) into the results and expand in
powers of the couplings.
\item
Using the resulting expressions evaluate
$v = F( G_c^{(2)}, G_c^{(3)}, m_0, g_0, f_0 )$ recursively by replacing
every occurrence of $v$ on the right hand side by the complete right
hand side and every occurrence of $G_c^{(2)}$ by the expression obtained
in the second step.
\end{itemize}
To every finite order in perturbation theory this straightforward
algebraic procedure stops after a finite number of steps since every
substitution is accompanied by additional factors of $g_0$ or $f_0$.

The necessary loop orders of the calculation are as follows.
Suppose you wish to calculate $\Gamma^{(n)}$ to $L$ loops.
Clearly the first part is to calculate all one-particle irreducible
$n$-point diagrams without tadpoles up to $L$ loops.
For the second part of the calculation one needs $v$ to $L'=L$ loops,
if $n= 2$ or 3.
For larger $n$ one needs $v$ only to $L'=L-1$ loops.
This calculation according to the procedure above requires all
$(L'-1)$-loop diagrams for $G_c^{(2)}(x,y)$ plus one additional loop
integration (for the coinciding arguments), as well as all $(L'-2)$-loop
diagrams for $G_c^{(3)}(x,y,z)$ plus two additional loop integrations.
%
\section{Conclusion}
With the help of Dyson-Schwinger equations we have discussed how tadpole
and snail diagrams occurring in the perturbative calculation of Green's
functions and vertex functions can be summed up by means of modified
Feynman rules.
The calculation has two parts.
The first one consists of the evaluation of Feynman diagrams without
tadpole and snail subdiagrams.
This reduces the number of diagrams significantly.
In the Feynman rules modified parameters are used.
In the second part the modified parameters are related to the original
ones.
This task is essentially algebraic and amounts to the iterative solution
of an equation for the tadpole.
\end{fmffile}


%

\begin{thebibliography}{99}
%
\bibitem{Kapusta}
J.\ Kapusta,
{\it Finite-temperature field theory},
Cambridge University Press 1989
%
\bibitem{BBH}
C.G.\ Boyd, D.E.\ Brahm, S.\ Hu,
Phys.\ Rev.\ \underline{D\,48} (1993) 4963
%
\bibitem{GKM}
C.\ Gutsfeld, J.\ K\"uster, G.\ M\"unster,
in preparation
%
\bibitem{Salam}
A.\ Salam,
Rev.\ Mod.\ Phys.\ \underline{33} (1961) 428
%
\bibitem{PC}
P.\ Cvitanovi\'{c},
{\it Field theory},
NORDITA lecture notes, January 1983
%
\bibitem{Rivers}
R.J.\ Rivers,
{\it Path integral methods in quantum field theory},
Cambridge University Press 1987
%
\bibitem{Dyson}
F.J.\ Dyson,
Phys.\ Rev.\ \underline{75} (1949) 1736
%
\bibitem{Schwinger}
J.\ Schwinger,
Proc.\ Nat.\ Acad.\ Sci.\ \underline{37} (1951) 452, 455
%
\bibitem{Symanzik}
K.\ Symanzik,
Z.\ Naturforschung \underline{9A} (1954) 809
%
\bibitem{Cardano}
G.\ Cardano, {\it Ars magna sive de regulis algebraicis}, Milano 1545
%
\end{thebibliography}
\end{document}